\documentclass[showpacs,preprint,12pt]{revtex4}

\usepackage[brazil, english]{babel}
\usepackage[utf8]{inputenc}
\usepackage{graphicx}
\usepackage{epsfig}
\usepackage{amssymb}
\usepackage{amsmath} 
\usepackage{color,soul}

\begin{document}

\title{Scalar and higher even spin glueball masses from an anomalous modified holographic model}
\author{Diego M. Rodrigues$^{1,}$}
\email[Eletronic address: ]{diegomr@if.ufrj.br}
\author{Eduardo Folco Capossoli$^{1,2,}$}
\email[Eletronic address:]{educapossoli@if.ufrj.br}
\author{Henrique Boschi-Filho$^{1,}$}
\email[Eletronic address: ]{boschi@if.ufrj.br}
\affiliation{$^1$Instituto de F\'{\i}sica, Universidade Federal do Rio de Janeiro, 21.941-972 - Rio de Janeiro-RJ - Brazil \\
 $^2$Departamento de F\'{\i}sica / Mestrado Profissional
em Práticas da Educação Básica (MPPEB),
 Col\'egio Pedro II, 20.921-903 - Rio de Janeiro-RJ - Brazil }

\begin{abstract}
In this work, within an anomalous modified holographic softwall model, we calculate analytically the masses of the scalar glueball with its radial excitations and higher even glueball spin states, with $P=C=+1$, from a single mass equation. Using this approach we achieved an unified treatment for both scalar and high even spin glueballs masses. 
Furthermore, we also obtain the Regge trajectory associated with the pomeron compatible with other approaches.
\end{abstract}


\maketitle


\section{Introduction}

In the late 1990s, Juan Maldacena's conjecture or AdS/CFT duality \cite{Maldacena:1997re, Gubser:1998bc, Witten:1998qj, Witten:1998zw, Aharony:1999ti} opened a myriad of possibilities in theoretical physics. This conjecture or duality  teaches us how to relate a superconformal Yang-Mills theory $({\cal N} = 4)$ with symmetry group $SU(N \to \infty)$ living in a flat four dimensional Minkowski space with a supergravity theory, which is the low energy limit of $IIB$ superstring theory living in a curved ten dimensional anti de Sitter space or $AdS_5 \times S^5$. After breaking the conformal symmetry one can build phenomenological models that approximately describe QCD. This approach, known as AdS/QCD, was implemented, for instance, in the hardwall model, where a hard cutoff is introduced at a certain value $z_{max}$ of the holographic coordinate, $z$, and a slice of the $AdS_5$ space is considered \cite{Polchinski:2001tt, Polchinski:2002jw, BoschiFilho:2002ta, BoschiFilho:2002vd}.

This approach proved to be very fruitful to investigate hadronic physics, such as, Regge trajectory studies. From the Regge  theory it can be shown for hadronic resonances that there is an approximate mathematical relationship between their squared masses $(m^2)$  and their spin quantum number $(J)$, given by:
\begin{equation}\label{regge}
J(m^2) \approx \alpha ' m^2  + \alpha_0\;,
\end{equation}
\noindent with $\alpha '$ and $\alpha_0$ constants.

This idea can be extended to bound states of gluons predicted by QCD, which have not yet been detected experimentally, called glueball states. Using the holographic hardwall model the masses for even glueball states and the Regge trajectories for both the pomeron and the odderon were obtained \cite{BoschiFilho:2005yh, Capossoli:2013kb, Rodrigues:2016cdb} 
which are compatible with those found in the literature. Besides, there are other recent works dealing with glueballs within other holographic approaches, see for instance \cite{Ballon-Bayona:2015wra, 
Brunner:2015yha, Parganlija:2016csi, Brunner:2015oqa, Brunner:2015oga}. For non-holographic approaches, see \cite{ Morningstar:1999rf, Meyer:2004jc, Meyer:2004gx, Chen:2005mg, Lucini:2001ej, Donnachie:1984xq, Donnachie:1985iz, Donnachie:2002en, Janowski:2011gt, Acharyya:2016fcn, Szczepaniak:2003mr, Mathieu:2008bf}. 

Another well known holographic AdS/QCD model is the softwall where a dilaton field is introduced playing the role of a soft cutoff. The main feature of this model is to produce linear Regge trajectories \cite{Karch:2006pv,Colangelo:2007pt}. As was discussed in \cite{BoschiFilho:2012xr, Li:2013oda, Capossoli:2015ywa, Capossoli:2016kcr, Capossoli:2016ydo} 
the Regge trajectories for glueballs coming from the original softwall model although linear are not in agreement with lattice data. In particular, in ref. \cite{Li:2013oda} it was proposed a dynamical modification of the softwall model for glueballs considering different dilaton fields. 
In the work \cite{Capossoli:2015ywa}  a modified holographic softwall was proposed based on its dynamical version but being analytically solvable. This model was used to calculate the masses of the scalar glueball states and its radial excitations as well as of higher even spin glueball states from a single mass equation. One problem of this approach is that one needs two different dilaton constant values for the scalar and high spin sectors. 

Here, in this work we overcome the above mentioned problem by introducing an anomalous dimension for the glueball operator, inspired by \cite{Gubser:2002tv, Braga:2014pxa}.  Then, we obtain from a single mass equation, with a unique dilaton constant value, the masses for the scalar and its radial excitations as well as for high even spin glueball states. The results achieved in this work are in agreement with other holographic and non-holographic approaches.

Even glueball states are related to the Reggeon whose intercept $\alpha_0$ in Eq. \eqref{regge} is close to unity, called the pomeron. In fact, in a $J \times m^2$ plane known as Chew-Frautschi plot the even glueball states lie on the Regge trajectory associated with the pomeron. The references \cite{Donnachie:1984xq, Donnachie:1985iz} present a well known Regge trajectory for the soft pomeron, given by:
\begin{equation}\label{land}
J(m^2) \approx 0.25 m^2  + 1.08\,,
\end{equation}
which will be used here for comparison with our results.

This work is organised as follows: In the section \ref{2},  we present a brief discussion about the dynamical softwall model. In the section  \ref{3} we introduce our anomalous modified softwall model and obtain a mass equation for the scalar and higher even spin glueballs. In section \ref{4}, we show our results for the glueball state masses and the Regge trajectory related to the pomeron. Finally, in section \ref {5}, we make our conclusions and last comments.


\section{The dynamical holographic softwall model}\label{2}

Let's start this section reviewing the dynamical softwall model discussed in \cite{Shock:2006gt, Capossoli:2016kcr, Capossoli:2016ydo, 
White:2007tu, Li:2013oda}. 
In this model the $5D$  dilaton-gravity action in the string frame is given by:
\begin{equation}\label{acao_corda}
S = \frac{1}{16 \pi G_5} \int d^5 x \sqrt{-g_s} \; e^{-2\Phi(z)} (R_s + 4 \partial_M \Phi \partial^M \Phi - V^s(\Phi))
\end{equation}
\noindent where $G_5$ is the Newton's constant and $g_s$ is the determinant of the metric tensor both in a $5-$dimensional space, $\Phi(z) = k z^2$ is the dilaton field, where $k \sim \Lambda^2_{QCD}$ and $V^s(\Phi)$ is the dilatonic potential. All of these parameters are in the string frame and the $5-$dimensional metric can be written as:
\begin{equation}\label{g_s}
ds^2 = g^s_{MN} dx^M dx^N = b^2_s(z)(dz^2 + \eta_{\mu \nu}dx^\mu dx^\nu); \; \; \;b_s(z) \equiv e^{A_s(z)}.
\end{equation}
The position vectors $x^M$ with capital latin indexes $M, N =0, 1, 2, 3, 4$ describe a 5D curved space while the vectors $x^\mu$ with greek indexes $\mu, \nu = 0, 1, 2,3$ describe the four dimensional Minkowski space with metric signature $\eta_{\mu \nu}={\rm diag} (-1, +1, +1, +1)$. The function $A_s(z)$ gives the metric warp factor and in the particular case where the 5D curved space is an $AdS_5$ it reads $A_s(z) = \log{\left( \frac{R}{z} \right)}$. Note that we are disregarding the $S^5$ subspace of the original AdS/CFT correspondence, as is usual in the AdS/QCD approach \cite{Karch:2006pv,Colangelo:2007pt,BoschiFilho:2012xr, Li:2013oda, Capossoli:2015ywa, Capossoli:2016kcr, Capossoli:2016ydo}. 

Although the string frame is advantageous to compute, for example, the vacuum expectation value of the loop operator it is more appropriate to use the Einstein frame once it simplifies the equations of motion. 
The metric tensors in both frames (string and Einstein) are related by a scaling transformation \cite{Li:2011hp}, such as: 
$ g^s_{MN} = g^E_{MN}e^{-2\Gamma}$. 
The dilaton potentials in these two frames are related by 
$ V^s = e^{2 \Gamma}V^E $ 
and for a particular choice $\Gamma = - \frac{2 \Phi}{3}$, it becomes 
$ V^s = e^{ \frac{-4 \Phi}{3}}V^E\, $.

Now one can write the $5D$  dilaton-gravity action \eqref{acao_corda} in the Einstein frame:
\begin{equation}\label{acao_einstein}
S = \frac{1}{16 \pi G_5} \int d^5 x \sqrt{-g_E} \; (R_E -\frac{4}{3} \partial_M \Phi \partial^M \Phi - V^E(\Phi)). 
\end{equation}

From this action one can obtain the equations of motion, given by:
\begin{equation}\label{eq_mov_e_1}
 E_{MN}  + \frac{1}{2}g^E_{MN}\left(\frac{4}{3} \partial_L\Phi \partial^L\Phi + V^E_G \right) - \frac{4}{3} \partial_M \Phi \partial^M \Phi = 0\;;
\end{equation}
\begin{equation}\label{eq_mov_e_2}
 \frac{8}{3 \sqrt{g_E}} \partial_M (\sqrt{g_E} \partial^M \Phi) - \partial_\Phi V^E(\Phi)  = 0\;,
\end{equation}
\noindent where $E_{MN}$ is the Einstein tensor.

Using the metric parametrisation given by (\ref{g_s}), the equations of the motion (\ref{eq_mov_e_1}) and (\ref{eq_mov_e_2}) can be written as:
\begin{equation}\label{eq_mov_e_2_1}
 -A''_E + A'^2_E - \frac{4}{9}\Phi'^2  = 0\;;
\end{equation}
\begin{equation}\label{eq_mov_e_2_2}
 \Phi'' + 3A'_E \Phi' - \frac{3}{8}e^{2A_E}\partial_\Phi V^E(\Phi) = 0\;,
\end{equation}
\noindent where we defined $\Phi'=\partial \Phi/ \partial z$, $A'=\partial A/ \partial z$ and
\begin{equation}\label{redef}
b_E (z) = b_s(z)e^{-\frac{2}{3}\Phi(z)} = e^{A_E(z)}\;; \qquad A_E(z) = A_s(z) - \frac{2}{3}\Phi(z)\;.
\end{equation}
\noindent Solving the equations (\ref{eq_mov_e_2_1}) and (\ref{eq_mov_e_2_2}) for the quadratic dilaton background, $\Phi(z)=kz^2$, one finds:
\begin{equation}\label{sol_eq_mov_e_2_1}
 A_E(z) = \log{\left( \frac{R}{z} \right)} - \log{\left(_0F_1\left(\frac 54, \frac{\Phi^2}{9}\right)\right)}\;, 
\end{equation}
\noindent and
\begin{equation}\label{sol_eq_mov_e_2_2}
 V^E(\Phi) = -\frac{12 ~ _0F_1(\frac14, \frac{\Phi^2}{9})^2}{R^2} + \frac{16 ~ _0F_1(\frac 54, \frac{\Phi^2}{9})^2\, \Phi^2}{3 R^2}\;,
\end{equation}
where $_0F_1(a,z)$ is the Kummer confluent hypergeometric function. 
Using (\ref{redef}) and (\ref{sol_eq_mov_e_2_1}), one can easily see that the warp factor in the string frame is
\begin{equation}\label{redef_2}
 A_s(z) = \log{\left( \frac{R}{z} \right)}  + \frac{2}{3}\Phi(z) - \log{\left[_0F_1\left(\frac 54, \frac{\Phi^2}{9}\right)\right]}\,, 
\end{equation}
\noindent which means that the metric (\ref{g_s}) is a deformed AdS space. Using $ V^s = e^{ \frac{-4 \Phi}{3}}V^E\, $, one has

\begin{equation}\label{vs}
 V^s(\Phi) =\exp\{-\frac 43 \Phi\} \left[ -\frac{12 ~ _0F_1(1/4, \frac{\Phi^2}{9})^2}{R^2} + \frac{16 ~ _0F_1(5/4, \frac{\Phi^2}{9})^2 \Phi^2}{3 R^2}\right]
\end{equation}

\noindent so that this potential generates the desired quadratic dilaton.

At this point, it is worth calling the attention for the replacement mentioned before. The warp factor in eq.  \eqref{redef_2} produces a modification in the $AdS_5$ space. We mean that this warp factor for the dynamical softwall model is not $AdS_5$ anymore, but for the UV limit or $z\rightarrow 0$, one has $A_s(z)|_{(z\rightarrow 0 )}\propto \log \left( \frac{R}{z} \right)$, meaning that the geometry is asymptotically $AdS$. 


\section{The anomalous modified holographic softwall model}\label{3}

After the discussion about dynamical softwall model, we will now describe the scalar glueball in 5D with the action in the string frame as in the original softwall model described in \cite{Karch:2006pv, Colangelo:2007pt}, and given by:
\begin{equation}\label{acao_soft1}
S =  \int d^5 x \sqrt{-g} e^{-\Phi(z)}\left[g^{MN} \partial_M{\cal G}\partial_N{\cal G} + M^2_{5} {\cal G}^2\right],
\end{equation}
\noindent but with the metric (\ref{g_s}) with the warp factor $A_s(z)$ given by \eqref{redef_2}.
 
 The corresponding equations of motion derived from this action are given by:
\begin{equation}\label{eom_1}
\partial_M[\sqrt{-g_s} \;  e^{-\Phi(z)} g^{MN} \partial_N {\cal G}] - \sqrt{-g_s} e^{-\Phi(z)} {{M}^2_{\rm 5}} {\cal G} = 0\,. 
\end{equation}
Now, using the ansatz $
{\cal G}(z, x^{\mu}) = v(z) e^{i q_{\mu} x^{\mu}}$  and defining $v(z) = \psi (z) e^{{B(z)} / {2}} $, with $ B(z) = \Phi(z) - 3A_s(z)$,  one gets a Schr\"odinger like equation: 
\begin{equation}\label{equ_5}
- \psi''(z) + \left[ \frac{B'^2(z)}{4}  - \frac{B''(z)}{2} + {M}^2_{\rm 5} \left( \frac{R}{z}\right)^2  e^{4kz^2/3} {\cal A}^{-2} \right] \psi(z) = - q^2 \psi(z)
\end{equation}
where ${\cal A}= _0F_1(5/4, {\Phi^2}/{9})$. This equation has no known analytical solution and was solved numerically in \cite{Li:2013oda, 
Capossoli:2016kcr, Capossoli:2016ydo}. 

Following \cite{Capossoli:2015ywa} we replace the warp factor $A_s(z)$  by: 
\begin{equation}\label{am}
{{A}}_M(z) = \log{\left( \frac{R}{z} \right)}  + \frac{2}{3}\Phi(z)\,.
\end{equation}
 Note that if this modified warp factor is replaced in Eqs. \eqref{eq_mov_e_2_1} and \eqref{eq_mov_e_2_2} then it would imply a non-quadratic dilaton profile. Besides, in this case one should solve a non-linear pair of differential equations which does not have an analytical solution. Actually, this modified warp factor is the starting point of the model we want to discuss here. 
 
Then, plugging \eqref{am}  in Eq. (\ref{equ_5}) one finds: 
\begin{equation}\label{equ_7}
- \psi''(z) + \left[ k^2 z^2 + \frac{15}{4z^2}  - 2k + {M}^2_{5} \left( \frac{R}{z}\right)^2  e^{4kz^2/3}\right] \psi(z) = (- q^2 )\psi(z).
\end{equation}

Expanding the exponential in the above equation we obtain at first order in $k$ 
\begin{equation}\label{equ_7_1_new}
- \psi''(z) + \left[ k^2 z^2 + \frac{15}{4z^2}  - 2k + {M}^2_{5} \left( \frac{R}{z}\right)^2   
+ \frac{4 kz^2}{3} {M}^2_{5} \left( \frac{R}{z}\right)^2\right] \psi(z) = (- q^2 )\psi(z)\,.
\end{equation}
The main reason to do this truncation is because potentials like \eqref{vs}  are confining as one can see in the left panels of figures 2 and 4 of ref. \cite{Capossoli:2016kcr}. So that effectively the coordinate z cannot grow indefinitely. Then, the IR behaviour is safe in the expansion considered here.
Also, as we will see in the next section, the parameter $k$ is small compared to other scales involved in this problem, such as the square of the glueball states' masses. 

Now, the equation \eqref{equ_7_1_new} is analytically solvable. From the eigenenergies  and associating $-q^ 2$ with the square of the masses of the 4D glueball states, one has:
\begin{equation}\label{adsw_1}
m_n^2 = \left[ 4n + 2\sqrt{4 + {M}^2_{5}R^2} + \frac{4}{3} {M}^2_{5}R^2 \right]k; \;\;\;\; (n=0, 1, 2, \cdots),
\end{equation}
\noindent where $n$ represents radial excitations of glueballs states, which means that $n= 0$ is the corresponding ground state.

Taking into account the AdS/CFT dictionary, it teaches us how to relate the five dimensional  mass of a scalar field in  the $AdS$ space and the conformal dimension $(\Delta)$ of an operator in the boundary theory: 
${M}^2_{5} R^2 = \Delta (\Delta - 4)\,$. For a massive tensor field with spin $J$ \cite{Karch:2006pv}, one has 
\begin{equation}\label{mass5}
{M}^2_{5} R^2 = \Delta (\Delta - 4) -J \,.
\end{equation}

In a super Yang-Mills theory defined on the boundary of the $AdS_5$ space, the scalar glueball $0^{++}$ is represented by the operator ${\cal O}_4 = Tr (F^2)\,$, which has conformal dimension $\Delta=4$ and ${M}_{5}  = 0$. 
In order to raise the spin, following \cite{deTeramond:2005su}, one can insert $J$ symmetrized covariant derivatives in a given operator with spin $S$ and then,
the total angular momentum after the insertion is $S+J$. Doing this for the scalar glueball state one has 
${\cal O}_{4+J} = FD_{\{\mu1 \cdots}D_{\mu J\}}F$,  with conformal dimension $\Delta = 4 + J$ and spin $J$.

Now, introducing the idea of an anomalous dimension ($\gamma$) \cite{Gubser:2002tv, Braga:2014pxa} for the glueball operator we can make the substitution in its conformal dimension  
\begin{equation}
 \Delta \to \Delta' = \Delta + \gamma (J)\,,
 \end{equation} 
 where $\Delta = 4 + J$. 
 Note that  we are only interested in the regime where the conformal dimension $\Delta$ for the scalar case obeys the condition  $2 < \Delta < 4$, which corresponds to deformations allowed by the Breitenlohner-Freedman (BF) bound \cite{Gubser:2008yx}. Note also that the UV behaviour does not match that of QCD asymptotic freedom. Of course, one can argue that this situation might need going beyond the supergravity approximation. Here, for our case we choose to match QCD at a finite scale, replacing asymptotic freedom by conformal symmetry, as considered in ref. \cite{Gubser:2008yx}. 
 
 Then, for a massive tensor field with spin $J$, using \eqref{mass5}, one has: 
\begin{equation}\label{mlinha}
 {M}^2_{5} R^2 = [4+J +  \gamma (J)]  [J+  \gamma (J) ] -J \,.
 \end{equation}

For sake of simplicity we are going to take a linear approximation for the anomalous dimension so that $\gamma (J)= \gamma_0 J$, where $\gamma_0$ is a dimensionless constant to be fixed later.  Note also that  in ref. \cite{Gubser:2002tv} they considered a more general function $\gamma (J)$, but in our case it suffices to take a linear behaviour to fit lattice data for the glueball masses and the Regge trajectory for the pomeron. 
Then, the 4D glueball masses given by Eq. \eqref{adsw_1} using Eq. \eqref{mlinha} can be read as 
\begin{eqnarray}\label{mold4}
m_n^2 &=& \Big[ 4n + 2\sqrt{4 +{(4+J+ \gamma_0 J)(J +\gamma_0 J) -J}} \cr
&& + \frac{4}{3}{(4+J+ \gamma_0 J)(J +\gamma_0 J) -J} \Big]\, k; \qquad\qquad 
(n=0, 1, 2, \cdots). 
\end{eqnarray} 
This equation represents the mass spectra of the even glueball states with spin $J$, which reduces to $ m_n^2 = \left[ 4n +4 \right] k \,$ for the scalar glueball case ($J=0$), in accordance with \cite{Capossoli:2015ywa}. 



\section{Even Spin Glueball States Spectra}\label{4}

In this section we will present our achievements related to the scalar glueball spectra (ground state and its radial excitations), and higher even spin glueball states masses from eq. \eqref{mold4}. From these results we will construct the corresponding Regge trajectory associated with the pomeron.

Before we show our results, we present here some values of the masses for the scalar glueball and its radial excitations (see Table \ref{tzero}) and other higher even spin glueball states  (see Table \ref{thigher}) obtained from lattice and other approaches. 


\begin{table}[h]

\centering
\begin{tabular}{|c|c|c|c|c|c|c|}
\hline 
 & $N_c = 3$ & \multicolumn{2}{c|}{$N_c = 3$ anisotropic lattice} & $N_c = 3$ & $N_c \rightarrow \infty$ & Average  \\ 
\hline 
$J^{PC}$  & ref. \cite{Meyer:2004gx} & ref. \cite{Morningstar:1999rf}  & ref. \cite{Chen:2005mg} & \multicolumn{2}{c|}{ref. \cite{Lucini:2001ej}} &  \\ 
\hline 
$0^{++}$ & 1.475(30)(65) &1.730(50)(80) &1.710(50)(80) & 1.58(11) & 1.48(07) & 1.595 \\ 
\hline 
$0^{++*}$ & 2.755(70)(120) & 2.670(180)(130) &  & 2.75(35) & 2.83(22) & 2.751\\ 
\hline 
$0^{++**}$ & 3.370(100)(150) &  &  &  &  & 3.370 \\ 
\hline 
$0^{++***}$ & 3.990(210)(180) &  &  &  &  & 3.990 \\ 
\hline 
\end{tabular} 
\parbox{4.7in}{\caption{\em Lightest scalar glueball and its radial excitation masses expressed in {\rm GeV}  from lattice. The numbers in the parenthesis represent the uncertainties. The last column represents mean values for each state.}\label{tzero}}
\end{table}

\begin{table}[h]

\centering
\begin{tabular}{|c|c|c|c|c|c|c|}
\hline 
 & $N_c = 3$ & \multicolumn{2}{c|}{$N_c = 3$ anisotropic lattice} & \multicolumn{2}{c|}{ Constituent models} & Average \\ 
\hline 
$J^{PC}$  & ref. \cite{Meyer:2004gx} & ref. \cite{Morningstar:1999rf}  & ref. \cite{Chen:2005mg} & ref. \cite{Szczepaniak:2003mr}  & ref. \cite{Mathieu:2008bf} & \\ 
\hline 
$2^{++}$ & 2.150(30)(100) &2.400(25)(120) &2.390(30)(120) & 2.42 & 2.59 & 2.39 \\ 
\hline 
$4^{++}$ & 3.640(90)(160) &  &  & 3.99 & 3.77 & 3.80 \\ 
\hline 
$6^{++}$ & 4.360(260)(200) &  &  &  & 4.60 & 4.48 \\ 
\hline 
\end{tabular} 
\parbox{4.7in}{\caption{\em Higher even spin glueball states masses expressed in {\rm GeV} from lattice and constituent model approaches. The numbers in the parenthesis represent the uncertainties related to lattice results. The last column represents mean values for each state.}\label{thigher}}

\end{table}

Our results for the glueball masses were obtained from the single equation \eqref{mold4}  and are shown in Table \ref{t1}. In order to achieve the best fit for the masses  we set the values  $k = 0.845$ GeV$^2$ and $\gamma_0 = -0.585$. 

In order to determine $k$ and $\gamma_0$ we have used a statistical analysis based on the r.m.s. error defined by:
\begin{equation}
 \delta_{rms}= \sqrt{ \frac 1{N-N_p} \sum_{i=1}^N \left( \frac {\delta O_i}{O_i} \right)^2} 
\end{equation}
where $N$ and $N_p$ are the number of measurements and parameters, respectively. Besides, $\delta O_i$ are the deviations between the data and the model prediction. Then, we minimised $\delta_{rms}$ finding the best values for $k$ and $\gamma_0$. Note that we have taken the average values shown in Tables \ref{tzero} and \ref{thigher} as input data.

 One can note that the value of the dilaton constant $k$ is small compared with any squared glueball masses including the lightest one, $m_{0^{++}}^2= 3.38$ GeV$^2$. Due to this, the expansion of the exponential presented in Eq. \eqref{equ_7} could be truncated at first order in $k$ meaningfully. Note also that the value $\gamma_0 = -0.585$ means that the dimension of the scalar glueball operator ${\cal O}_4=Tr (F^2)$ is effectively equal to  $\Delta'= 3.415$. The higher spin glueballs have effective conformal dimension $\Delta'= 3.415+J$. Note that this result is in qualitative agreement with \cite{Gubser:2008yx} where the conformal dimension of the glueball operator decreases with the anomalous dimension.

\begin{table}[h]
\vspace{0.5 cm}
\centering
\begin{tabular}{|c|c|c|c|c|c|c|c|c|c|c|}
\hline
 &  \multicolumn{9}{c|}{Glueball States $J^{PC}$}  \\  \cline{2-10}
 &$0^{++}$& $0^{++\ast}$& $0^{++\ast\ast}$& $0^{++\ast\ast\ast}$ & $2^{++} $ & $4^{++}$ & $6^{++}$ & $8^{++}$ & $10^{++}$   \\
\hline \hline
Masses (GeV)                                   
&\, 1.84 \,&\, 2.60 \,&\, 3.18 \, &\, 3.67 \,&\, 2.64 \,&\, 3.52 \,& \, 4.42 \, &\, 5.31 \,&\, 6.21 \, 
\\ \hline 
Deviations (\%)
 &\, 15  \,&\, 5.4 \,&\, 5.6 \, &\, 7.7 \,&\, 10.4 \,&\, 7.3 \,& \, 1.4 \, &\, \hl{} \,&\, \hl{} \, 
\\ \hline 
\end{tabular}
\parbox{4.8in}{\caption{\em Masses and deviations for the glueball states $J^{PC}$ with even $J$, including the scalar glueball and its radial excitations, from the anomalous modified softwall model using the Eq. (\ref{mold4}) with $k= 0.845$ {\rm GeV}$^2$ and $\gamma_0=-0.585$.}\label{t1}}
\end{table}

The total r.m.s. errors for the scalar and high spin sectors are respectively found to be  10.9 \% and 9.1\%, which show good agreement between the model predictions and the data considered. 

\begin{figure}[h] 
  \centering
  \includegraphics[scale = 0.5, angle=0]{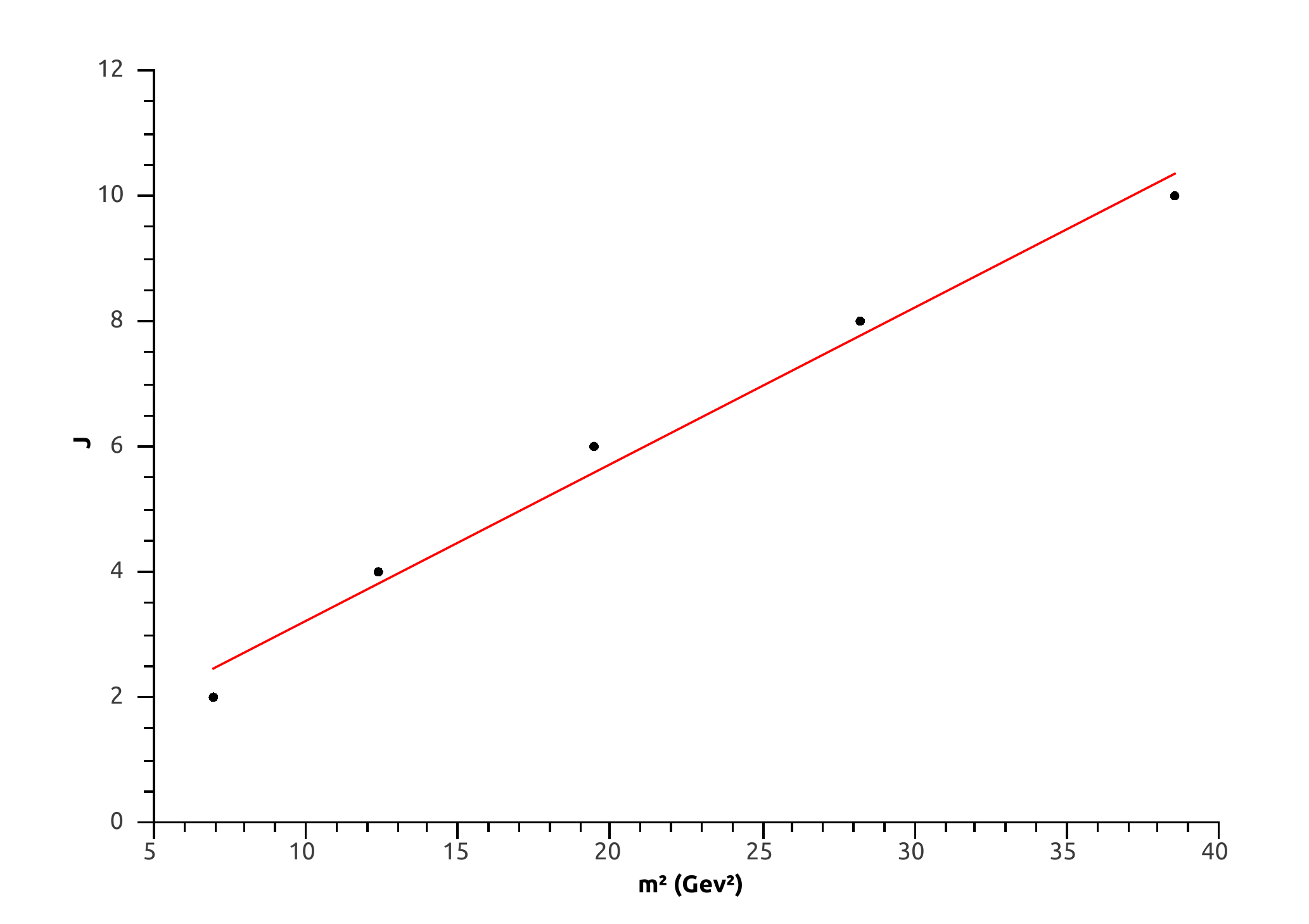} 
 \parbox{4.7in} {\caption{\em Approximate Regge trajectory for the pomeron using data from table \ref{t1}, for the states $2^{++} $, $4^{++}$,  $6^{++}$,  $8^{++}$ and $10^{++}$,  from the anomalous modified softwall model using Eq. (\ref{mold4}) with  $k= 0.845$ {\rm GeV}$^2$ and $\gamma_0=-0.585$.}\label{plot}}
\end{figure}

Using the values found in Table \ref{t1}, for the states $2^{++} $, $4^{++}$,  $6^{++}$,  $8^{++}$ and $10^{++}$, we got the following Regge trajectory for the pomeron: 
\begin{equation}\label{R2}
J(m^2) = (0.25 \pm 0.02) m^2  + (0.73 \pm 0.42)\,, 
\end{equation} 
 in accordance with Eq. \eqref{land}. The errors in this equation come from a regular linear regression method applied to  Table \ref{t1} data. 
The plot related to the Regge trajectory for the pomeron from Eq. \eqref{R2}, using the data found in table \ref{t1}, is presented in the Figure \ref{plot}.

One should note that only the original softwall model 
provides linear Regge trajectories among the holographic models. However, such model does not give good results for glueball masses. Here in our modified and anomalous softwall model the Regge trajectories are no longer linear although the results for the glueball masses are very good with r.m.s. error less than 11\%. Any way, we can find an approximate linear Regge trajectory for the even glueball states.


\section{conclusions}\label{5}

In this work, we have proposed an anomalous modified softwall model (which is analytically solvable) in order to unify the spectra of the scalar and higher even spin glueball states with just one dilaton constant value ($k=0.845\, {\rm GeV}^2$). 
Note that in the previous work \cite{Capossoli:2015ywa} a single mass equation for the scalar and higher even spin glueball states was obtained but to fit lattice data two different values of the dilaton constant $k$ were needed. Here, with the introduction of an anomalous dimension in the conformal dimension of the glueball operators, this problem is overcome. 

Note that in ref.  \cite{Gubser:2002tv} the anomalous dimension for high spin fields is a logarithm function of spin $J$.
Here, we just used a linear function of the spin $J$ as an approximation for lower spins, presenting good results for our model in comparison with lattice data and other models. 

Besides, the data obtained here from our model and presented in table \ref{t1} provided a Regge trajectory for the pomeron in agreement with the literature  \cite{Donnachie:1984xq, Donnachie:1985iz}. 

Finally, it is important to mention that the spectrum obtained here in particular for the scalar glueball and its radial excitations fits better the lattice data summarised in Table \ref{tzero} than the ones presented, for instance, in ref. \cite{Capossoli:2015ywa, Capossoli:2016kcr, Capossoli:2016ydo}.  In these references, the calculated mass for the scalar glueball $0^{++}$ was too low in comparison with the literature.

\begin{acknowledgments} We would like to acknowledge useful conversations with Nelson R. F. Braga. 
D.M.R. is supported by  Conselho Nacional de Desenvolvimento Cient\'\i fico e Tecnol\'ogico (CNPq), E.F.C. is partially supported by PROPGPEC-Colégio Pedro II, and H.B.-F. is partially supported by CNPq. 
 
 \end{acknowledgments}


\end{document}